\documentclass[12pt,a4paper,fleqn]{article}
\usepackage{amsmath,amssymb,amsfonts,amsthm,amsopn,graphicx}
\usepackage{tikz} 
\usepackage{mathrsfs} 
\usetikzlibrary{arrows}
\usepackage{hyperref}

\usepackage[T1,T2A]{fontenc}
\allowdisplaybreaks[3]
\numberwithin{equation}{section} 
\newtheorem{theorem}{Theorem}[section]

\newtheorem{stat}[theorem]{Statement}

\newcommand{\ds}{\displaystyle}
\def\EXP{\textrm{{\large e}}}
\newcommand{\ii}{{\sf i}}

\newcommand{\hl}{^1\!\!/\!_2}

\newcommand{\Ch}{\textrm{\scalebox{1.2}{$\chi$}}}

\newcommand{\Si}{\textrm{\scalebox{1.2}{$\sigma$}}}

\def\uop{\boldsymbol{u}}

\def\vop{\boldsymbol{v}}
\def\kop{\boldsymbol{k}}
\def\eop{\boldsymbol{e}}
\def\fop{\boldsymbol{f}}

\def\Clebsh{\textrm{\scalebox{1.5}{$\kappa$}}}

\usepackage[body={15.5cm,22cm}]{geometry}
\begin{document}

\title{Functional Bethe Ansatz for a $\sinh$-Gordon model with real $q$.}
\author{Sergey Sergeev
\thanks{ 
Department of Fundamental and Theoretical Physics,
         Research School of Physics and Engineering,
    Australian National University, Canberra, ACT 0200, Australia}
\thanks{    
    Faculty of Science and Technology, 
   University of Canberra, Bruce ACT 2617, Australia
}}

\date{}

\maketitle

\begin{abstract}
Recently, Bazhanov and Sergeev have described an Ising-type 
integrable model which can be identified as a $\sinh$-Gordon-type model with an infinite number of states but with a 
real parameter $q$. This model is the subject of Sklyanin's Functional Bethe Ansatz. We develop in this paper the whole technique of the FBA which includes:
\begin{itemize}
\item[1.] Construction of eigenstates of an off-diagonal element of a monodromy matrix.  Most important ingredients of these eigenstates are the Clebsh-Gordan coefficients of the corresponding representation.
\item[2.] Separately, we discuss the Clebsh-Gordan coefficients, as well as the Wigner's 6j symbols, in details. The later are rather well known in the theory of $3D$ indices.
\end{itemize}
Thus, the Sklyanin basis of the quantum separation of variables is constructed. The matrix elements of an eigenstate of the auxiliary transfer matrix in this basis are products of functions satisfying the Baxter equation. Such functions are called usually the $Q$-operators. We investigate the Baxter equation and $Q$-operators from two points of view.
\begin{itemize}
\item[3.] In the model considered the most convenient 
Bethe-type variables are the zeros of a Wronskian of two well defined particular solutions of the Baxter equation. 
This approach works perfectly in the thermodynamic limit. We calculate the distribution of these roots in the thermodynamic limit, and so we reproduce in this way the partition function of the model.
\item[4.] The real parameter $q$, which is the standard quantum group parameter, plays the role of the absolute temperature in the model considered. Expansion with respect to $q$ (tropical expansion) gives an alternative way to establish the structure of the eigenstates. In this way we classify the elementary excitations over the ground state.
\end{itemize}

\end{abstract}


\tableofcontents


\section{Introduction}

In this paper we apply Sklyanin' method of the Functional Bethe Ansatz  \cite{Sklyanin} 
to the recently formulated Ising-type model related to a simple Weyl algebra representation of the $\mathscr{U}_q(\hat{sl}_2)$ quantum algebra \cite{BS}.  This is in fact a very well known representation, usually referred to as a ``$\sinh$-Gordon'' one, usually it was studied in the framework of either cyclic representation (Fateev-Zamolodchikov model, or the Chiral Potts model), or in the framework of the Faddeev's modular double (Faddeev-Volkov model). The case we consider here corresponds to the simple Weyl algebra with $|q|<1$, however it also can be solved by the Functional Bethe Ansatz method as all models mentioned above.

We construct the Sklyanin basis of the ``quantum separation of variables'' where the matrix elements of an eigenstate is the product of solutions of Baxter equation (Baxter's $\hat{Q}$ operator). The Baxter equation is later studied in two approaches. The first one is based on a special form of Bethe Ansatz, where the Bethe variables are zeros of some special Wronskian of two well defined ``unphysical'' solutions of the Baxter equation. In this way we derive a functional equation for the density of these zeros for the ground state, and thus we reproduce the partition function of the model. The second approach is based on a so-called ``tropical expansion''. This approach allows one to classify the excitations over the ground state.

The paper is organised in a lapidary way. In the rest of this section we remind the method of the Functional Bethe Ansatz, fix the notations for the representation of $\mathscr{U}_q(\hat{sl}_2)$ we use, and fix some definitions of several convenient special functions. 

Then, in Section 2, we formulate the solution of Sklyanin's problem of diagonalisation of the off-diagonal element $\hat{B}(\lambda)$ of the auxiliary monodromy matrix. This diagonalisation is based on a properly defined Clebsh-Gordan-Wigner functions appeared in the theory of the invariants of three dimensional manifolds. We devote a special Section 3 to details and to a proof of statements given in Section 2. The structure of the expressions is close to those obtained by Kharchev and Lebedev for the Toda chain and the modular double \cite{KL}.

The first (Bethe Ansatz) approach to the Baxter equation can be found in Section 4. And, finally, we discuss the excitations in Section 5 using a ``tropical expansion'' approach.

\subsection{The Functional Bethe Ansatz Method}

Here we'd like to remind the Reader the Sklyanin method to construct the eigenstates for an auxiliary transfer matrix for a class of $\mathscr{U}_q(\hat{sl}_2)$ models.

Everywhere in this paper we use the standard notations:
\begin{equation}
[x]\;\stackrel{def}{=}\; x-x^{-1}\;,\quad (x;q^2)_\infty \;\stackrel{def}{=}\; \prod_{k=0}^\infty (1-q^{2k}x)\;.
\end{equation}

The starting point for the FBA is the intertwining of an auxiliary $L$-operators with the help of the six-vertex $R$-matrix,
\begin{equation}
R(\lambda) \;=\; 
\left(\begin{array}{cccc}
[q\lambda] & 0 & 0 & 0 \\
0 & [\lambda] & [q] & 0 \\
0 & [q] & [\lambda] & 0 \\
0 & 0 & 0 & [q\lambda]
\end{array}
\right)
\end{equation}
Then, using a standard $\mathscr{U}_q(\hat{sl}_2)$ Lax operator,
\begin{equation}\label{QG}
L(\lambda)\;=\;
\left(\begin{array}{cc}
[\lambda\kop] & \fop \\
\\
\eop & [\lambda\kop^{-1}]
\end{array}
\right)\;,\quad
\kop\eop=q\eop\kop,
\quad
\kop\fop=q^{-1}\fop\kop,
\quad
[\eop,\fop] = [q][\kop^2]\;,
\end{equation}
satisfying the $RLL$ intertwining relation
\begin{equation}\label{RLL}
R(\frac{\lambda}{\mu}) L(\lambda)\otimes L(\mu) \;=\;
(1\otimes L(\mu)) (L(\lambda) \otimes 1) R(\frac{\lambda}{\mu})\;,
\end{equation}
one obtains the monodromy matrix,
\begin{equation}\label{mon}
M_n(\lambda) \;=\;
L_1(\lambda)L_2(\lambda)\cdots L_n(\lambda) \;=\;
\left(
\begin{array}{cc}
\ds \hat{A}_n(\lambda) & \ds \hat{B}_n(\lambda)\\
\\
\ds \hat{C}_n(\lambda) & \ds \hat{D}_n(\lambda)
\end{array}
\right)\;.
\end{equation}
Here and in what follows $n$ stand for the chain length. The Sklyanin method is based on only two relations for the matrix elements of the monodromy matrix following from (\ref{RLL}):
\begin{equation}\label{AD}
\begin{array}{l}
\ds [q\frac{\lambda}{\mu}] \hat{A}_n(\lambda) \hat{B}_n(\mu) \;=\;
[\frac{\lambda}{\mu}] \hat{B}_n(\mu) \hat{A}_n(\lambda) + [q] \hat{A}_n(\mu) \hat{B}_n(\lambda)\;,\\
\\
\ds [q^{-1}\frac{\lambda}{\mu}] \hat{D}_n(\lambda) \hat{B}_n(\mu) \;=\;
[\frac{\lambda}{\mu}] \hat{B}_n(\mu) \hat{D}_n(\lambda) - [q] \hat{D}_n(\mu) \hat{B}_n(\lambda)\;.
\end{array}
\end{equation}
Let
\begin{equation}
| z \rangle \;=\; | z_1,z_2,\dots,z_{n-1}\rangle 
\end{equation}
be an eigenstate for the operator $\hat{B}_n(\lambda)$: 
\begin{equation}
\hat{B}_n(\lambda) | z \rangle \;=\; | z \rangle \, \prod_{j=1}^{n-1} [\frac{\lambda}{z_j}]\;.
\end{equation}
Then the relation (\ref{AD}) leads to
\begin{equation}\label{AD2}
\hat{A}_n(z_j) |z\rangle \;=\; a(z_j) |\dots qz_j\dots\rangle\;,\quad
\hat{D}_n(z_j) |z\rangle \;=\; d(z_j) |\dots q^{-1}z_j\dots\rangle
\end{equation}
with some coefficients $a(z)$ and $d(z)$. 
Thus, an eigenstate of the auxiliary transfer matrix $\langle \Psi|$,
\begin{equation}
\langle \Psi| (\hat{A}_n(\lambda) +\hat{D}_n(\lambda))\;=\; T(\lambda) \langle \Psi |\;,
\end{equation}
being considered in the basis of the eigenstates $|z\rangle$ of the operator $\hat{B}_n(\lambda)$, has a remarkably simple form,
\begin{equation}\label{BA}
\langle \Psi| z \rangle \;=\;\prod_{j=1}^{n-1} Q(z_j)
\end{equation}
where $Q(z)$ is the solution of Baxter's  $TQ$ equation
\begin{equation}\label{TQ}
T(z) Q(z) \;=\; a(z) Q(qz) + d(z) Q(q^{-1}z)\;.
\end{equation}
A thoughtful Reader may observe that the chain length is $n$, while the number of parameters in the state $|z\rangle$ is one less. The reason is that there is an analogue of ``total spin'' operator in the spectrum of the auxiliary transfer matrix,
\begin{equation}
\boldsymbol{K}\;=\;\kop_1\kop_2\cdots \kop_n\;,
\end{equation}
and its eigenvalue is an additional hidden parameter for all eigenstates.

Thus, the Sklyanin method of the Functional Bethe Ansatz implies three steps:
\begin{equation}\label{Skl}
\begin{array}{ll}
\ds (1) & \textrm{To construct the eigenstates of the off-diagonal operator $\hat{B}_n(\lambda)$.}\\
\ds (2) & \textrm{To find the coefficients $a(z)$ and $d(z)$ from equations (\ref{AD2}).}\\
\ds (3) & \textrm{To solve the Baxter equation (\ref{TQ}).}
\end{array}
\end{equation}

\subsection{The algebra and the representation}

Let $\uop,\vop$ be the simple Weyl algebra,
\begin{equation}
\uop\vop\;=\;q\vop\uop\;.
\end{equation}
We use the following representation of simple Weyl algebra:
\begin{equation}\label{rep1}
\langle \varphi | \psi \rangle \;=\; \psi(v)\;,\quad v\;=\;\EXP^{\ii\varphi}\;,
\end{equation}
then
\begin{equation}\label{rep2}
\langle \varphi | \vop | \psi \rangle \;=\; v \psi(v)\;,\quad
\langle \varphi | \uop | \psi \rangle \;=\; \psi(qv)\;.
\end{equation}
Note that the operator $\uop$ becomes diagonal in the Fourier basis,
\begin{equation}\label{uqm}
\langle \varphi | \psi_m\rangle \;=\; v^m\quad \Rightarrow\quad
\uop  |\psi_m\rangle \;=\; |\psi_m\rangle \, q^m\;.
\end{equation}

Let us choose now the following representation $\pi$ of the algebra $\mathscr{U}_q(sl_2)$,
\begin{equation}
\pi(\kop)\;=\;\vop\;,\quad \pi(\fop)\;=\;\uop\;,\quad 
\pi(\eop)\;=\;\mathtt{E}(s,\vop) \uop^{-1}\;,
\end{equation}
where
\begin{equation}\label{E}
\mathtt{E}(s,v)\;=\;[v/s][q/sv]\;=\; \frac{q}{s^2}+\frac{s^2}{q} - \frac{q}{v^2}-\frac{v^2}{q}\;.
\end{equation}
One usually mentions the Casimir operator at this point.
\begin{equation}
\textrm{Casimir}\;\stackrel{def}{=}\;\eop\fop+q\kop^{-2}+q^{-1}\kop^2\;=\;qs^{-2}+q^{-1}s^2\;,
\end{equation}
in order to emphasise that $s$ is the representation parameter. Thus we study the auxiliary Lax operator of the form
\begin{equation}
L(\lambda)\;=\;\left(\begin{array}{cc}
[\lambda\vop] & \uop \\
\\
\mathtt{E}(s,\vop) \uop^{-1} & [\lambda\vop^{-1}]
\end{array}
\right)\;.
\end{equation}
Here we ``undress'' the operator $\fop$ taking $\pi(\fop)\;=\;\uop$ since the first object of our interest is the off-diagonal operator $\hat{B}_n(\lambda)$. The Reader may identify this $L$-operator with the class of so-called $\sinh$-Gorodon $L$ operators \cite{FV}.

\subsection{The main special function and the main identity}

Concluding the Introduction, we'd like to define our main special function and the identities which it satisfies. Namely, let
\begin{equation}\label{Si}
\Si(v) \;=\; \frac{(-q/v;q)_\infty}{(v;q)_\infty}\;,\quad \frac{\Si(qv)}{\Si(v)}\;=\;-[v]\;,\quad
\frac{1}{\Si(v)}\;=\;\Si(-q/v)\;.
\end{equation}
Garoufalidis and Kashaev mentioned the importance of this function in \cite{Kash2}. It satisfies the fundamental Spiridonov-type identity known as the ``beta-integral for the pentagon equation'' from one hand side, and also known as the ``star-triangle relation'' for the model considered from the other hand side \cite{BS}:
\begin{equation}\label{PT}
\frac{1}{2\pi\ii\kappa_s} \oint \; \frac{dv}{v} \;
\prod_{j=1}^3 \frac{\Si(a_jv)}{\Si(b_jv)}\;=\;
\prod_{i,k=1}^3 \Si(\frac{b_j}{a_k})^{-1}\;,\quad
\frac{b_1b_2b_3}{a_1a_2a_3}\;=\;q^2\;,
\end{equation}
where the integration contour could be chosen so that 
\begin{equation}
\max\left( |q/b_j| \right)\,<\,|v|\,<\,\min \left( 1/|a_j| \right)\;.
\end{equation}
As well, here for the shortness
\begin{equation}
\kappa_s\;=\;\frac{(-q;q)_\infty}{(q;q)_\infty}\;.
\end{equation}

One more useful function is
\begin{equation}\label{Cl}
\Clebsh(v_1,v_2)\;=\;\frac{\Si(v_1)\Si(v_2)}{\Si(v_1v_2)}\;.
\end{equation}
Garoufalidis and Kashaev called this function ``an index for an ideal tetrahedron'' \cite{DGG,Kash2}.
Below we will see that it is a 3j symbol, i.e. a CG coefficient. The ``beta-integral'' plays the role of the Biedenharn-Elliot identity for $6j$ symbols. The identity (\ref{PT}) can be re-written as
\begin{equation}\label{6j}
\frac{1}{2\pi\ii\kappa_s}
\oint \frac{dx'}{x'}
\Clebsh(x'v_1,\frac{v_2}{x'}) \Clebsh(yv_1v_2,\frac{x'}{y}v_3) 
\Clebsh(\frac{y}{xx'},xx') \;=\;
\Clebsh(\frac{y}{x}v_1,\frac{v_2v_3}{y})
\Clebsh(xv_2,\frac{v_3}{x})\;.
\end{equation}
One more useful identity is the inversion relation for the GC coefficients,
\begin{equation}\label{inv}
\frac{1}{2\pi\ii\kappa_s} \oint \frac{dx'}{x'} 
\Clebsh(\frac{y}{xx'},xx') 
\Clebsh(\frac{x'x''}{y},\frac{1}{x'x''}) \;=\;
2\pi\kappa_s \delta(\varphi_x-\varphi_{x''})\;,\quad
x\;=\;\EXP^{\ii\varphi_{x}}\;.
\end{equation}
Finally, we define one more convenient function,
\begin{equation}\label{J}
J(s,v)\;=\;\frac{\Si(sv)}{\Si(s/v)}\;,\quad
\frac{J(s,v)}{J(s,q^{-1}v)}\;=\;\mathtt{E}(s,v)\;,\quad
\frac{J(s,qv)}{J(s,v)}\;=\;\mathtt{E}(s,v^{-1})\;,
\end{equation}
where $\mathtt{E}$ is defined by  (\ref{E}).

\section{The Eigenstates for $\hat{B}_n(\lambda)$}

An answer to the first two items of Sklyanin's program (\ref{Skl}) is formulated in this Section. The final answer is rather lengthy and cumbersome. So, we commence with a set of definitions.
\\

\subsection{Definitions and notations}

Let us introduce the following variables,
\begin{equation}
\mu_{i,j}\;,\quad i\;=\;1,\dots,n-1\;,\quad j\;=\;1,\dots,i
\end{equation}
and 
\begin{equation}
v_1\;=\;\EXP^{\ii\varphi_1}\;,\quad v_2\;=\;\EXP^{\ii\varphi_2}\;,\quad\dots\;,
v_n\;=\;\EXP^{\ii\varphi_n}\;.
\end{equation}
Next, we define the following uncomfortable and clumsy system of notations,
\begin{equation}\label{tbl}
\begin{array}{ll}
\boldsymbol{\mu}_1\;=\;\mu_{1,1}\;, & v^{(1)}\;=\;v_1\;,\\
\\
\boldsymbol{\mu}_2\;=\;\mu_{2,1}\mu_{2,2}\;, & v^{(2)}\;=\;v_1v_2\;,\\
\\
\boldsymbol{\mu}_3\;=\;-\mu_{3,1}\mu_{3,2}\mu_{3,3}\;, & v^{(3)}\;=\;v_1v_2v_3\;,\\
\\
\vdots & \vdots \\
\\
\boldsymbol{\mu}_{n-1}\;=\;\epsilon_{n}\mu_{n-1,1}\mu_{n-1,2}\cdots \mu_{n-1,n-1}\;, &
v^{(n-1)}\;=\;v_1v_2\cdots v_{n-1}\;.
\end{array}
\end{equation}
The signs appeared somewhere here are defined by 
\begin{equation}\label{eps}
\epsilon_2^{}\;=\;\epsilon_2'\;=\;1\;, \quad
\epsilon_{n+1}^{}\;=\;\epsilon_n'\;,\quad \epsilon_{n+1}'\;=\;(-)^{n+1}\epsilon_n^{}\;.
\end{equation}
A meaning of this sequence with the period four will become clear later.

Нext, we introduce the vector notations: top down for the table (\ref{tbl}),
\begin{equation}
\boldsymbol{\mu}\;=\;\{\boldsymbol{\mu}_1,\, \boldsymbol{\mu}_2,\, \dots\, ,\, \boldsymbol{\mu}_{n-1}\}
\end{equation}
and from left to right,
\begin{equation}
\mu_{\ell}\;=\;\{\mu_{\ell,1},\, \mu_{\ell,2},\, \dots\, , \,\mu_{\ell,\ell}\}\;,\quad
v\;=\;\{v_1,\,v_2,\,\dots\,,v_n\}
\end{equation}
The Clebsh-Gordan coefficients are defined above (\ref{Cl}), however we need also a ``long CG coefficient from left to right'',
\begin{equation}\label{LCl}
\Clebsh_n(\boldsymbol{\mu})\;=\;
\prod_{\ell=1}^{n-1} \Clebsh(\boldsymbol{\mu}_\ell v^{(\ell)},\frac{\boldsymbol{\mu}_{\ell-1}}{\boldsymbol{\mu}_\ell}v_{\ell+1})\;,
\end{equation}
where formally $\boldsymbol{\mu}_0=1$. Next, the kernel for the eigenstate of $\hat{B}_n(\lambda)$ is given by
\begin{equation}\label{ker}
\mathcal{F}_n(\mu) \;=\; 
\left(
\prod_{i=1}^{n-2} \prod_{j=1}^i J(s,\mu_{i,j})
\right)
\times
\left(
\prod_{k=2}^{n-2} \prod_{1\leq i < j \leq k} [\frac{\mu_{k,i}}{\mu_{k,j}}]
\right)
\times
\left(
\prod_{k=1}^{n-2} \prod_{i=1}^{k+1} \prod_{j=1}^k \Si(\frac{\mu_{k+1,i}}{\mu_{k,j}})
\right)
\end{equation}
Here $J(s,v)$ are defined by (\ref{J}). The Reader can find below few examples of this scary expression (for the sake of shortness we temporarily re-define  $\mu_1=x$, $\mu_2=y$, $\mu_3=z$):
\begin{equation}
\begin{array}{l}
\ds \mathcal{F}_2\;=\;1\;,\\
\\
\ds \mathcal{F}_3(\mu)\;=\; J(s,x)
\Si(\frac{y_1}{x}) \Si(\frac{y_2}{x})\;,\\
\\
\ds \mathcal{F}_4(\mu)\;=\; \mathcal{F}_3(\mu)\; [\frac{y_1}{y_2}]\;
J(s,y_1) J(s,y_2) \;
\Si(\frac{z_1}{y_1})
\Si(\frac{z_2}{y_1})
\Si(\frac{z_3}{y_1})
\Si(\frac{z_1}{y_2})
\Si(\frac{z_2}{y_2})
\Si(\frac{z_3}{y_2})
\end{array}
\end{equation}
In particular, 
\begin{equation}\label{fnn}
\frac{\mathcal{F}_{n+1}(\mu)}{\mathcal{F}_n(\mu)}\;=\;
\left(
\prod_{j=1}^{n-1} J(s,\mu_{n-1,j}) 
\right)
\times
\left( \prod_{1\leq i < j \leq n-1} [\frac{\mu_{n-1,i}}{\mu_{n-1,j}}]
\right)
\times
\left(
\prod_{i=1}^n
\prod_{j=1}^{n-1}
\Si(\frac{\mu_{n,i}}{\mu_{n-1,j}})
\right)\;.
\end{equation}
Finally, the last indigestible definition in this section:
\begin{equation}\label{main}
\Omega_n(v,\mu_{n-1})\;=\;
\left(
\prod_{i=1}^{n-2}
\prod_{j=1}^i 
\oint 
\frac{1}{2\pi\ii\kappa_s} \frac{d\mu_{i,j}}{\mu_{i,j}}
\right)\times
\Clebsh_n(v,\boldsymbol{\mu})\;
\mathcal{F}_n(\mu)
\end{equation}
There are a lot of integrals in all formulas above. The integrals are well defined as the integrals over the unit circle when 
\begin{equation}
|q|<|s|<1
\end{equation}
and with the extra convention 
\begin{equation}
\Si(v)\;\to\; \Si(v\EXP^{-\epsilon})
\end{equation}
in the case when a pole of $\Si$ lies on the unit circle.
\\

Now we are ready to define the state 
\begin{equation}\label{thestate}
\langle \varphi | z \rangle \;=\; \Omega_n(v;z)\;,
\end{equation}
where\footnote{Often we will write shortly $\Omega_n(v;x)=\Omega_n(z)$. As well, taking into account the Dirac notations, we will imply $\uop_k \Omega_n(v;z)= \Omega_n(\dots qv_k\dots; z)$.} for shortness $z\;=\;\mu_{n-1}$, i.e.
\begin{equation}
\{z_1,z_2,\dots,z_{n-1}\} \;=\; \{\mu_{n-1,1},\mu_{n-1,2},\dots,\mu_{n-1,n-1}\}\;,
\end{equation}
and
\begin{equation}
\varphi\;=\;\{\varphi_1,\varphi_2,\dots,\varphi_{n}\}\;,\quad v\;=\;\{\EXP^{\ii\varphi_1},\EXP^{\ii\varphi_2},\dots,\EXP^{\ii\varphi_n}\}\;.
\end{equation}

\subsection{The main statement}

The following statement takes place:
\begin{stat}\label{Stat1}
The state $|z\rangle$ defined by (\ref{thestate},\ref{main},...) satisfies
\begin{equation}\label{B}
\hat{B}_n(\lambda) |z\rangle \;=\; 
|z\rangle \; \epsilon_n^{}\, \prod_{j=1}^{n-1} [\frac{\lambda}{z_{j}}]\;.
\end{equation}
In addition 
\begin{equation}\label{A}
\hat{A}_n(z_j) | z \rangle \;=\; 
| \dots q z_j\dots\rangle \, (-)^n\,\epsilon_{n-1}\, \mathtt{E}(s,z_j^{-1})
\end{equation}
and 
\begin{equation}\label{D}
\hat{D}_n(z_j) | z \rangle \;=\; 
| \dots q^{-1}z_j\dots\rangle \, \epsilon_{n-1}^{}\, \mathtt{E}(s,z_j)^{n-1}
\end{equation}
where the set of ``epsilons'' is defined by  (\ref{eps}).
\end{stat}
\noindent
\textbf{Proof} of this statement -- the whole next section.

\section{Proof of the Statement \ref{Stat1}}

For the proof of the Statement \ref{Stat1} we will use the mathematical induction.

\subsection{$n=2$.} As the base, we take $n=2$ and consider explicit form of $\hat{B}_2(\lambda)$:
\begin{equation}\label{b2}
\hat{B}_2(\lambda)\;=\; [\lambda\vop_1^{}] \uop_2^{} + \uop_1^{} [\lambda\vop_2^{-1}] \;.
\end{equation}
Taking now
\begin{equation}
\langle \varphi_1,\varphi_2| \Omega\rangle \;=\; \Omega(v_1,v_2)
\end{equation}
where notations (\ref{rep1},\ref{rep2}) are implied and $\Omega(v_1,v_2)$ is unknown, we obtain equation for the eigenstate:
\begin{equation}
\langle \varphi_1,\varphi_2| \hat{B}_2(\lambda) | \Omega \rangle \;=\;
[\lambda v_1] \Omega (v_1,qv_2) + [\lambda/v_2] \Omega (qv_1,v_2) \;=\;  q^m [\frac{\lambda}{x}] \Omega(v_1,v_2)\;.
\end{equation}
This equation comprises two independent equations corresponding to the order $\lambda$ and to the order $\lambda^{-1}$, so that the solution is unique:
\begin{equation}
\frac{\Omega(qv_1,v_2)}{\Omega(v_1,v_2)}\;=\; q^m \frac{[xv_1]}{[v_1v_2]}\;,\quad
\frac{\Omega(v_1,qv_2)}{\Omega(v_1,v_2)}\;=\; q^m \frac{[v_2/x]}{[v_1v_2]}\;,
\end{equation}
Checking this with (\ref{uqm},\ref{Si},\ref{Cl}), we can fix
\begin{equation}
\Omega(v_1,v_2)\;=\;\Clebsh(xv_1,v_2/x) (v_1v_2)^m\;.
\end{equation}
All similar expressions with CG coefficients will contain simple factors like $(v_1v_2)^m$, $(v_1v_2v_3)^m$, and so on, so for shortness we will omit such factor putting $m=0$. This corresponds to the ground state.
\\

Expression (\ref{b2}) justifies the term ``Clebsh-Gordan coefficient'' since $\hat{B}_2(\lambda) \;=\; \lambda \Delta^{(-)}(\eop) - \lambda^{-1}\Delta^{(+)}(\eop)$ has the structure of a co-product, see below.
\\

So, taking $m=0$, we obtain
\begin{equation}
\langle \varphi_1,\varphi_2|x\rangle\;=\;
\Omega_2(x) \;=\; \Clebsh(xv_1,\frac{v_2}{x})\;,\quad
\hat{B}_2(\lambda)|x\rangle \;=\; |x\rangle \, [\frac{\lambda}{x}]\;.
\end{equation}
Now, consider $\hat{A}_2(\lambda)$ and $\hat{D}_2(\lambda)$:
\begin{equation}
\hat{A}_2(\lambda)\;=\;[\lambda\vop_1^{}][\lambda\vop_2^{}]+\mathtt{E}(s,\vop_2^{})\uop_1^{}\uop_2^{-1}\;,\quad
\hat{D}_2(\lambda)\;=\;[\lambda\vop_1^{-1}][\lambda\vop_2^{-1}]+\mathtt{E}(s,\vop_1^{})\uop_1^{-1}\uop_2^{}\;.
\end{equation}
The direct verification gives
\begin{equation}
\hat{A}_2(x) |x\rangle \;=\; |qx\rangle \mathtt{E}(s,x^{-1})\;,\quad
\hat{D}_2(x) |x\rangle \;=\; |q^{-1}x\rangle \mathtt{E}(s,x)\;.
\end{equation}
Thus the base for the mathematical induction is satisfied.

\subsection{Clebsh-Gordan coefficients and Wigner $6j$ symbols}

Before we continue the proof of the Statement \ref{Stat1}, let us discuss the general construction of the Clebsh-Gordan coefficients.

To do it, consider two co-products:
\begin{equation}
\Delta_n^{(-)}(\fop)\;=\;\sum_{j=1}^n \vop_1^{}\cdots \vop_{j-1}^{}\uop_j^{}\vop_{j+1}^{-1}\cdots \vop_n^{-1}\;
\end{equation}
and
\begin{equation}
\Delta_n^{(+)}(\fop)\;=\;\sum_{j=1}^n \vop_1^{-1}\cdots \vop_{j-1}^{-1}\uop_j^{}\vop_{j+1}^{}\cdots \vop_n^{}\;.
\end{equation}
Note,
\begin{equation}
\hat{B}_n(\lambda)\;=\;\lambda^{n-1} \Delta_n^{(-)}(\fop)\;+\;\cdots\;+\;(-\lambda)^{-n+1}\Delta_n^{(+)}(\fop)\;.
\end{equation}
``Long CG coefficients'' appear as the basis for an Eigenspace of these co-products. 
A list of ``long CG coefficients from left to right'' is the following:
\begin{equation}
\begin{array}{l}
\ds \Clebsh_2(x)\;=\; \Clebsh(xv_1,\frac{v_2}{x})\;,\\
\\
\ds \Clebsh_3(x,y)\;=\;\Clebsh(xv_1,\frac{v_2}{x}) \Clebsh(yv_1v_2,\frac{x}{y}v_3)\;,\\
\\
\ds \Clebsh_4(x,y,z)\;=\;\Clebsh(xv_1,\frac{v_2}{x}) \Clebsh(yv_1v_2,\frac{x}{y}v_3)
\Clebsh(zv_1v_2v_3,\frac{y}{z}v_4)\;,\\
\\
\ds \vdots\\
\\
\ds \Clebsh_n(\mu_1,\dots,\mu_{n-1}) \;=\; 
\prod_{j=1}^{n-1} \Clebsh(\mu_j v_1\cdots v_j,\frac{\mu_{j-1}}{\mu_j} v_{j+1})\;.
\end{array}
\end{equation}
The last expression is a copy of (\ref{LCl}).
For all these ``long CG coefficients'' we have 
\begin{stat}\label{stat2}
\begin{equation}
\Delta_n^{(\pm)}(\fop) \Clebsh(\mu_1,\dots,\mu_{n-1}) \;=\; \Clebsh(\mu_1,\cdots,\mu_{n-1}) \mu_{n-1}^{\pm 1}\quad
\forall\;\mu_1,\dots,\mu_{n-2}\;.
\end{equation}
where bra-s and ket-s are implied but omitted for the shortness.
\end{stat}
\noindent
\textbf{Proof:} is straightforward and simple. \hfill $\square$
\\

However, there exists also a set of ``long CG coefficients from right to left'':
\begin{equation}
\begin{array}{l}
\ds \Clebsh_2'(x)\;=\; \Clebsh(xv_1,\frac{v_2}{x})\;,\\
\\
\ds \Clebsh_3'(x,y)\;=\;\Clebsh(\frac{y}{x}v_1,\frac{v_2v_3}{y}) \Clebsh(xv_2,\frac{v_3}{x})\;,\\
\\
\ds \Clebsh_4'(x,y,z)\;=\;\Clebsh(\frac{z}{y}v_1,\frac{v_2v_3v_4}{z}) \Clebsh(\frac{y}{x}v_2,\frac{v_3v_4}{y})
\Clebsh(x v_3,\frac{v_4}{x})\;,\\
\\
\ds \vdots\\
\\
\ds \Clebsh_n'(\mu_1,\dots,\mu_{n-1}) \;=\; 
\prod_{j=1}^{n-1} \Clebsh(\frac{\mu_{n-j}}{\mu_{n-j-1}} v_j, \frac{v_{j+1}\cdots v_n}{\mu_{n-j}})\;.
\end{array}
\end{equation}
They also satisfy the Statement \ref{stat2}. Therefore, there exists a linear operator transforming one basis to the other.

One can use the completeness relation in order to obtain a kernel of this linear operator. In two sites it reads 
\footnote{We use here the following normalisation for the measure and for the delta-functions,
$$
\kappa_s \sum_m \to 2\pi\kappa_s \delta(\varphi-\varphi')\;,\quad
\frac{1}{2\pi\kappa_s} \frac{dv}{v} \to \frac{1}{\kappa_s} \delta_{m,m'}
$$
}:
\begin{equation}
\begin{array}{l}
\ds \kappa_s \sum_m \frac{1}{2\pi\ii\kappa_s} \oint \frac{dx}{x} \Clebsh(xv_1,\frac{v_2}{x}) 
\Clebsh(\frac{1}{x v_1'},\frac{x}{v_2'}) \left(\frac{v_1v_2}{v_1'v_2'}\right)^m\;=\;
(2\pi\kappa_s)^2 \delta(\varphi_1^{}-\varphi_1')\delta(\varphi_2^{}-\varphi_2')\;,\\
\\
\ds \frac{1}{(2\pi\ii\kappa_s)^2}
\oint\oint \frac{dv_1}{v_1}\frac{dv_2}{v_2} \Clebsh(xv_1,\frac{v_2}{x}) \Clebsh(\frac{1}{x'v_1},\frac{x'}{v_2})
(v_1v_2)^{m-m'}\;=\;2\pi \delta(\varphi_{x}-\varphi_{x'}) \delta_{m,m'}\;.
\end{array}
\end{equation}
These formulas are related to the inversion (\ref{inv}).

Applying the completeness relations, one can obtain the intertwiner for $\Clebsh_3^{}(x,y)$ 
and $\Clebsh_3'(x',y)$. The final form of the intertwining relation reads
\begin{equation}\label{K33}
\begin{array}{l}
\ds \frac{1}{2\pi\ii\kappa_s} \oint \frac{dx}{x} \Clebsh_3^{}(x,y) \Clebsh(\frac{y}{xx'},xx')\;=\;
\Clebsh_3'(x',y)\;,\\
\\
\ds 
\frac{1}{2\pi\ii\kappa_s} \oint \frac{dx'}{x'} \Clebsh_3'(x',y) \Clebsh(\frac{xx'}{y},\frac{1}{xx'})\;=\;
\Clebsh_3^{}(x,y)\;.
\end{array}
\end{equation}
This linear operator is by definition the Wigner $6j$ symbol which apparently coincide with the $3j$ symbol, and the first relation of (\ref{K33}) is given in the Introduction, equation (\ref{6j}).

\subsection{Proof of the Statement \ref{Stat1} continued. The induction step}

Suppose now, the Statement \ref{Stat1} is true for some $n$. Namely, there is defined a function 
\begin{equation}
\Omega_n(z)\;,
\end{equation}
constructed with the help of definitions (\ref{LCl},\ref{ker}), including the ``long CG coefficient'', and where for shortness
\begin{equation}
z\;=\;\{z_1,z_2,\dots,z_{n-1}\}\;=\;\{\mu_{n-1,1},\mu_{n-1,2},\dots,\mu_{n-1,n-1}\}\;.
\end{equation}
and where
\begin{equation}
\boldsymbol{z}\;=\;\epsilon_n\, z_1z_2\cdots z_{n-1}\;,
\end{equation}
and which satisfies 
\begin{equation}\label{BAn}
\hat{B}_n(\lambda) \Omega_n(z) \;=\; \Omega_n(z)\, \epsilon_n^{}\,  \prod_{j=1}^{n-1} [\frac{\lambda}{z_j}]\;,\quad
\hat{A}_n(z_j) \Omega_n(z) \;=\; \Omega( \dots qz_j\dots)\, \epsilon_n' \, \mathtt{E}(s,z_j^{-1})\;.
\end{equation}
Consider next the case $n+1$, when the monodromy matrix has the form
\begin{equation}\label{n1}
M_{n+1}\;=\; \left(
\begin{array}{cc}
\ds 
\hat{A}_n(\lambda)[\lambda\vop_{n+1}^{}] + \mathtt{E}(s,\vop_{n+1}^{}) \hat{B}_n(\lambda) \uop_{n+1}^{-1}
&
\ds \hat{A}_n(\lambda) \uop_{n+1}^{}+\hat{B}_n(\lambda) [\lambda\vop_{n+1}^{-1}]\\
\\
\cdots & \ds \hat{C}_n(\lambda) \uop_{n+1}^{} + \hat{D}_n(\lambda) [\lambda\vop_{n+1}^{-1}]
\end{array}
\right)
\end{equation}
so that 
\begin{equation}
\hat{B}_{n+1}(\lambda)\;=\;\hat{A}_n(\lambda) \uop_{n+1}^{}+\hat{B}_n(\lambda) [\lambda\vop_{n+1}^{-1}]\;.
\end{equation}

Consider now the action of $\hat{B}_{n+1}(\lambda)$ on the state 
\begin{equation}
\Psi(z;\boldsymbol{\mu})\;=\;\Omega_n(z) \Clebsh(\boldsymbol{\mu} v^{(n)},\frac{\boldsymbol{z}}{\boldsymbol{\mu}}v_{n+1})
\end{equation}
where the additional CG coefficient brings the previous ``long CG coefficient'', already included into $\Omega_n$, to the required length. The scalar parameter $\boldsymbol{\mu}$ is yet undefined. 
The straightforward consequence of (\ref{BAn}) is 
\begin{equation}
\hat{B}_{n+1}(z_j) \Psi(z;\boldsymbol{\mu})\;=\; \epsilon_n'\, \mathtt{E}(s,z_j^{-1})\,\Psi(\dots qz_j\dots;\boldsymbol{\mu})\;.
\end{equation}
Therefore, it follows
\begin{stat}
\begin{equation}\label{f4}
\begin{array}{ll}
\ds \ds \hat{B}_{n+1}(\lambda) \Psi(z;\boldsymbol{\mu}) & \ds =\;\epsilon_n^{} 
[\frac{\lambda \boldsymbol{z}}{\boldsymbol{\mu}}] \, \prod_{j=1}^{n-1} [\frac{\lambda}{z_j}] 
\, \Psi(z;\boldsymbol{\mu})\\
\\
&\ds 
+ \; \epsilon_n'
\sum_{j=1}^{n-1} \Psi(\dots qz_j\dots;\boldsymbol{\mu}) \mathtt{E}(s,z_j^{-1})
\prod_{k\neq j} \frac{\ds [\frac{\lambda}{z_k}] }{\ds [\frac{z_j}{z_k}] }\;.
\end{array}
\end{equation}
\end{stat}
\noindent\textbf{Proof:} this formula is an identity for the polynomials in $\lambda$ of degree $n$. It is satisfied in $n-1$ points $\lambda=z_j$, and also it has consistent asymptotic when $\lambda\to 0,\infty$ (see the statement \ref{stat2}) . These are the sufficient requirements. \hfill $\blacksquare$

We take next (yet undefined) function 
\begin{equation}
\mathcal{G}_n(z;\mu)\;,\quad \mu\;=\;\{\mu_1,\mu_2,\dots,\mu_n\}\;,\quad \boldsymbol{\mu}=
\epsilon_{n+1}^{}\prod_j \mu_j\;,
\end{equation}
and try 
\begin{equation}
\Omega_{n+1}(\mu) \;=\; \left(\prod_{j=1}^{n-1} \oint \frac{dz_j}{z_j}\right)
\Psi(z;\boldsymbol{\mu}) \mathcal{G}(z;\mu)\;.
\end{equation} 
In this case the desired eigenstate equation 
\begin{equation}
B_{n+1}(\lambda) \Omega_{n+1}(\mu) \;=\; \Omega_{n+1}(\mu) \; \epsilon_{n+1}^{}\,\prod_{\ell=1}^n [\frac{\lambda}{\mu_\ell}]
\end{equation}
together with (\ref{f4}) gives
\begin{equation}\label{f5}
\ds \epsilon_n^{}
[\frac{\lambda \boldsymbol{z}}{\boldsymbol{\mu}}] \, \prod_{j=1}^{n-1} [\frac{\lambda}{z_j}] +
\epsilon_n'
\sum_{j=1}^{n-1} \mathtt{E}(s,qz_j^{-1})
\prod_{k\neq j} \frac{\ds [\frac{\lambda}{z_k}] }{\ds [q^{-1}\frac{z_j}{z_k}] }
\frac{\mathcal{G}_n(\dots q^{-1}z_j\dots;\mu)}{\mathcal{G}_n(z;\mu)}\;=\;\epsilon_{n+1}^{}
\prod_{j=\ell}^n [\frac{\lambda}{\mu_\ell}]\;.
\end{equation}
Leading in $\lambda$ terms here are cancelled. Taking $\lambda=z_j$ in this equation, we obtain 
\begin{equation}
\frac{
\mathcal{G}_n(\dots q^{-1}z_j\dots;\mu)
}
{\mathcal{G}_n(z;\mu)}\;=\;\epsilon_n'\epsilon_{n+1}^{}
\frac{1}{\mathtt{E}(s,qz_j^{-1})}\times
\left(
\prod_{k\neq j} \frac{\ds [q^{-1}\frac{z_j}{z_k}] }{\ds [\frac{z_j}{z_k}] }
\right)
\times
\prod_{\ell=1}^n [\frac{z_j}{\mu_\ell}]
\end{equation}
so that 
\begin{equation}\label{g2}
\mathcal{G}_n(z;\mu) \;=\;
\left(
\prod_{j=1}^{n-1}
J(s,z_j)
\right)
\times
\left(
\prod_{1\leq i<j\leq n-1} [\frac{z_i}{z_j}] 
\right)
\times
\left(
\prod_{\ell=1}^{n}
\prod_{j=1}^{n-1} 
\Si(\frac{\mu_\ell}{z_j})
\right)\;,
\end{equation}
if $\epsilon_{n+1}^{}=\epsilon_n'$, what is one of the conditions (\ref{eps}).

Comparing this expression with (\ref{fnn}), one can verify their equivalence upon the identification 
$z=\mu_{n-1}$ and $\mu=\mu_n$, so that the expression (\ref{ker}) is reproduced. Thus, the formula (\ref{B}) from the Statement \ref{Stat1} is proven. \hfill $\blacksquare$
\\

The proof of the equation (\ref{A}) is trivial. It is enough to take the formula (\ref{n1}) and to derive after simple computations 
\begin{equation}
\hat{A}_{n+1}(\lambda) \;=\;
[\lambda \vop_{n+1}^{}]\uop_{n+1}^{-1} \hat{B}_{n+1}(\lambda) + \mathtt{E}(s,\lambda^{-1}) \uop_{n+1}^{-1} \hat{B}_n(\lambda)\;,
\end{equation}
so that (\ref{A}) is provided instantly, and since 
\begin{equation}
\uop_{n+1}^{-1}\hat{B}_n(\mu_\ell) \Psi(z;\boldsymbol{\mu}) \;=\; 
\Psi(x;\boldsymbol{\mu}) \epsilon_n \prod_{j=1}^{n-1} [\frac{\mu_\ell}{z_j}]\;=\;
(-)^{n-1}\epsilon_n \Psi(z;q\boldsymbol{\mu}) \frac{\mathcal{G}(z;\dots q\mu_\ell \dots)}{\mathcal{G}(z;\mu)}\;,
\end{equation}
then
\begin{equation}
\hat{A}_{n+1}(\mu_\ell) \Omega_{n+1}(\mu) \;=\;
(-)^{n-1}\epsilon_{n} \; \mathtt{E}(s,\mu_\ell^{-1}) \; \Omega_{n+1}(\dots q\mu_\ell \dots)\;,
\end{equation}
and thus the condition $\epsilon_{n+1}'\;=\;(-)^{n-1}\epsilon_n^{}$, cf. (\ref{eps}), is fixed. \hfill $\blacksquare$ $\blacksquare$
\\

Finally in this Section, we turn to the proof of the formula (\ref{D}). It could be proven as follows. Instead of ``the long CG coefficients from left to right'', one have to consider ``the long CG coefficients from right to left''. This means that one has to take the formula (\ref{main}) and to start to push the CG coefficients from right to left with the help of the relation (\ref{K33}), using the identity (\ref{PT}) on each step. This computation in details is too lengthy. Here we demonstrate this procedure for $n=3$. The initial formula (\ref{main}) for $n=3$ reads
\begin{equation}\label{o3}
\Omega_3(y_1,y_2) \;=\;
\frac{1}{2\pi\ii\kappa_s} \oint \frac{dx}{x} 
\Clebsh(xv_1,\frac{v_2}{x}) 
\Clebsh(y_1y_2v_1v_2,\frac{x}{y_1y_2}v_3)
\frac{\Si(sx)}{\Si(s/x)} \Si(\frac{y_1}{x})\Si(\frac{y_2}{x})\;.
\end{equation}
The same $\Omega_3(y_1,y_2)$, rewritten ``from right to left'', becomes
\begin{equation}\label{op3}
\begin{array}{l}
\ds \Omega_3(y_1,y_2) \;=\; 
\frac{\Si(sy_1)\Si(sy_2)}{\Si(s/y_1)\Si(s/y_2)}\times\\
\\
\ds \frac{1}{2\pi\ii\kappa_s}
\oint \frac{dx}{x}
\Clebsh(\frac{y_1y_2}{x}v_1,\frac{v_2v_3}{y_1y_2})
\Clebsh(xv_2,\frac{v_3}{x})
\frac{\Si(s/x)}{\Si(sx)} \Si(\frac{x}{y_1})\Si(\frac{x}{y_2})\;.
\end{array}
\end{equation}
The structure of the integral here corresponds to the construction of the Eigenvector $| z \rangle$ using the similar mathematical induction but ``from right to left'' rather then ``from left to right''\footnote{
Just for reference, the induction ``from right to left'' reads 
$$
\begin{array}{l}
\ds \hat{B}_{n+1}(\lambda)\;=\;[\lambda\vop_1] \hat{B}_n(\lambda) + \uop_1 \hat{D}_n(\lambda)\;,\\
\\
\ds \hat{D}_{n+1}(\lambda) \;=\; \mathtt{E}(s,\lambda) \uop_1^{-1} \hat{B}_n(\lambda) + [\lambda\vop_1^{-1}]\uop_1^{-1} \hat{B}_{n+1}(\lambda)\;,\\
\\
\ds \Psi_n'(z;\boldsymbol{\mu})\;=\;\Clebsh(\frac{\boldsymbol{\mu}}{\boldsymbol{z}} v_1^{},\frac{\vop^{(1)}}{\boldsymbol{\mu}})\Omega_n'(z)\;.
\end{array}
$$
As the consequence, $\ds \hat{D}_{n+1}(\mu_\ell) \Omega_{n+1}(\mu)\;=\; \epsilon_n\;\mathtt{E}(s,\mu_\ell)^{n-1}\;\Omega_{n+1}(\dots q^{-1}\mu_\ell\dots )$.
}. However, this computation provides an extra multiplier, which is responsible for the power $n-1$ of $\mathtt{E}$ in the action of $\hat{D}_n$, see (\ref{D}).
\hfill $\blacksquare$ $\blacksquare$ $\blacksquare$
\\

An important conclusion to be made from the relations(\ref{o3}) and (\ref{op3}) is the following. 
The function $\Omega_n(z)$ is by construction analytical for $|z_j|\leq 1$. One can make the analytical continuation 
\begin{equation}\label{anal}
\Omega_n(z) \;=\; \left(\prod_{j=1}^{n-1} J(s,z_j)\right)^{n-2} \Omega_n'(z)\;,
\end{equation}
this transformation is in particular given by (\ref{o3},\ref{op3}), so that $\Omega'_n(z)$ is analytical for  $|z_j|\geq 1$. This observation is important for the next section.

\section{Baxter equation}

The separated basis $|z_1,\dots,z_{n-1}\rangle$ of the Eigenstates of operator $\hat{B}_n(\lambda)$ has been constructed in the previous Sections, so that the problem of the diagonalization of the auxiliary transfer matrix is reduced by means of (\ref{BA}) to the solution of the Baxter equation.

\subsection{Formulation of the problem}

So, gathering together (\ref{TQ}) and (\ref{A},\ref{D}), we come to the following equation for the component of the wave function (\ref{BA}), 
\begin{equation}\label{TQ1}
T(z) Q(z) \;=\; \epsilon_{n-1}^{} \left( (-)^n \mathtt{E}(s,z^{-1}) Q(qz) + \mathtt{E}(s,z)^{n-1} Q(q^{-1}z) \right)\;.
\end{equation}
where $z$ now is a scalar complex variable, and $\epsilon_{4n}\;=\;-1$, otherwise $\epsilon_n\;=\;1$. This is a strange prediction. The sign $(-)^n$ in (\ref{TQ1}) is good, but the common multiplier $\epsilon_{n-1}$, which equals to $-1$ for the chain lengths  $5,9,13,\dots$, looks strange. We do not have explanation to this. In what follows, we accept everywhere $\epsilon_{n-1}=1$.

Next we need an intermediate step. Equation (\ref{TQ1}) is not symmetric, one can bring it to a symmetric form introducing\footnote{This transformation is indeed the symmetrisation since the same equation for an inhomogeneous chain looks as follows:
$$
T(z) Q(z) \;=\;  (-)^n \mathtt{E}(s_n,z^{-1}) Q(qz) + \left(\prod_{k=1}^{n-1} \mathtt{E}(s_k,z)\right) Q(q^{-1}z)\;.
$$
 }
\begin{equation}\label{Qp}
Q'(z)\;=\;J(s,z) Q(z)\;,
\end{equation}
so that the Baxter equation becomes
\begin{equation}\label{TQp}
T(z) Q'(z) \;=\; (-)^n Q'(qz) + \mathtt{E}(s,z)^n Q'(q^{-1}z)\;.
\end{equation}
Now we can turn to an analytical structure of $Q'(z)$. Function $Q'(z)$, as well as the function $\Omega_n(z)$, is analytical for $|z|\leq 1$. Expression (\ref{anal}) gives the analytical continuation of $\Omega_n(z)$ to $|z|\geq 1$. During this continuation, it seems, some zeros may disappear, so the change (\ref{Qp}) seems to be reasonable. As the result, we postulate that the poles of $Q'(z)$ are given by 
\begin{equation}\label{qh1}
Q'(z)\;=\;\frac{H(z)}{(sz,-qz/s;q)_\infty^{n}}\;,
\end{equation}
where $H(z)$ is an entire function. Then the Baxter equation for the entire $H$ becomes
\begin{equation}\label{TH}
\ds T(z) H(z)  \; =\; (-)^n
(1-sz)^{n}(1+q\frac{z}{s})^{n}  H(qz) \;+\; 
(1-\frac{s}{z})^{n}(1+\frac{q}{sz})^{n} H(q^{-1}z)\;.
\end{equation}

\subsection{Comparison with previous results}

It could be interesting to compare the Baxter equation (\ref{TQ1}) arisen in this paper with the same equation from \cite{BS,S23} and obtained in a completely different way.

The result from \cite{BS,S23} reads: the Baxter equation for the \emph{operator} $\hat{Q}$ becomes (after transformation of notations to the current ones\footnote{
To be explicit, the $L$- and $Q$- operators from \cite{S23},
\begin{equation}\label{LQexplicit}
L=\left(\begin{array}{cc}
\ds [\frac{xx'}{\lambda^2}\vop] & \ds \uop [\frac{x'}{x}\vop]\\
\ds \uop^{-1}[\frac{x'}{x}\vop^{-1}] & \ds [\frac{xx'}{\lambda^2}\vop^{-1}]
\end{array}
\right)\;,\quad
\langle a|\hat{Q}(\lambda)|b\rangle=\prod_i \frac{W_{x/\lambda}(a_i-b_i) W_{\lambda/x'}(a_{i+1}-b_i)}{W_{x/x'}(b_{i+1}-b_i)}
\end{equation}
where
$$
W_x(n)\;=\;x^n \frac{(-q^{1+n}x^2;q^2)_\infty}{(-q^{1+n}/x^2;q^2)_\infty}\;,
$$
give
$$
T(\lambda)\hat{Q}(\lambda)=[\frac{x^{\prime 2}}{\lambda^2}]^n\hat{Q}(q^{1/2}\lambda) + [\frac{x^2}{\lambda^2}]^n \hat{Q}(q^{-1/2}\lambda)\;,
$$
what corresponds to (\ref{TQ3}) with the identification 
$\ds z=\frac{xx'}{\lambda^2}$, $\ds s=\frac{x}{x'}$,
$\ds \frac{x^{\prime 2}}{\lambda^2}=\frac{z}{s}$, $\ds \frac{x^2}{\lambda^2}=sz$.
}
\begin{equation}\label{TQ3}
T(z) \hat{Q}(z) \;=\; [sz]^n \hat{Q}(qz) + [z/s]^n \hat{Q}(q^{-1}z)\;.
\end{equation}
The \emph{operator} $\hat{Q}(z)$, satisfying (\ref{TQ3}), has the following set of poles according to \cite{BS,S23}:
\begin{equation}\label{qh2}
\hat{Q}(z) \;=\; \frac{H(z)}{(-q/sz,-qz/s;q)_\infty^n}\;,
\end{equation}
where $H(z)$ -- entire. 

One can verify straightforwardly, function $H(z)$ from (\ref{qh2}) satisfies the same equation (\ref{TH}) as required. 

Note, the \emph{operator} $\hat{Q}(z)$ satisfies also the Wronskian relation,
\begin{equation}\label{WrQ}
\hat{Q}(q^{-1}z) \hat{Q}(-z) - \hat{Q}(z) \hat{Q}(-q^{-1}z)\;=\;\textrm{const} \times \left(
\left(\frac{\Si(-z/s)}{\Si(-sz)}\right)^n - \left(\frac{\Si(z/s)}{\Si(sz)}\right)^n\right)\;,
\end{equation}
where $\textrm{const}=\kappa_s^{-2N}$ in the normalisation (\ref{LQexplicit}).

\subsection{The method to solve the Baxter equation for entire $H(z)$}

The following statement is true:
\begin{stat}\label{stat3}
The necessary and sufficient condition for the quantisation of the spectrum of $T(z)$ is the entireness of function $H(z)$ in the Baxter equation (\ref{TH}).
\end{stat}
\noindent
We will formulate the method of solution of (\ref{TH}) as the \textbf{proof} of this statement. The method we use is closely related to the results obtained in e.g. \cite{GW,FVS,RRS}. 

Ergo, we consider equation
\begin{equation}\label{TQ4}
T(z) H(z) \;=\; (-)^n a(z) H(qz) + d(z) H(q^{-1}z)\;,
\end{equation}
where 
\begin{equation}
a(z)\;=\;(1-sz)^n(1+q\frac{z}{s})^n\;,\quad d(z)\;=\; 
(1-\frac{s}{z})^n(1+\frac{q}{sz})^n\;,
\end{equation}
and, what is essential,
\begin{equation}\label{T2}
T(z)\;=\; 2z^n + \cdots + 2(-z)^{-n}\;,
\end{equation}
where the ground state condition 
\begin{equation}
\vop_1\vop_2\cdots\vop_n\;=\;1
\end{equation} 
is taken into account. Since equation (\ref{TQ4}) is the second order difference equation, we can talk about a two-dimensional basis of its solution space. We suggest the following particular basis:
\begin{equation}\label{Hpm}
H_{\pm}(z)\;=\;F_{\pm}(z) \;\Ch_{\pm}(z)\;,
\end{equation}
where
\begin{equation}
\frac{F_{+}(z)}{F_{+}(q^{-1}z)}\;=\;(-z)^n d(z)\;,
\end{equation}
so that
\begin{equation}\label{Fp}
F_{+}(z)\;=\;\frac{\ds (\frac{s}{z},-\frac{q}{sz};q)_\infty^n}{V_{+}(z)}\;,\quad
\frac{V_{+}(q^{-1}z)}{V_{+}(z)}\;=\;(-z)^n\;,
\end{equation}
and where $V_{+}(z)$ is some theta–function, yet undefined. The resulting equation for $\Ch_{+}(z)$ becomes
\begin{equation}\label{TCh1}
(-z)^n T(z) \Ch_+(z)\;=\; \Ch_{+}(q^{-1}z) + (1-s^2z^2)^n(1-q^2\frac{z^2}{s^2})^n \Ch_{+}(qz)\;.
\end{equation}
In its turn,
\begin{equation}
\frac{F_{-}(z)}{F_{-}(qz)}\;=\; (-z)^{-n} a(z)\;,
\end{equation}
so that 
\begin{equation}\label{Fm}
F_{-}(z)\;=\;
\frac{\ds (sz,-q\frac{z}{s};q)_\infty^n}{V_{-}(z)}\;,\quad
\frac{V_{-}(z)}{V_{-}(qz)}\;=\;(-z)^n\;,
\end{equation}
and the resulting equation for $\Ch_{-}$ becomes
\begin{equation}\label{TCh2}
z^{-n}T(z)\Ch_{-}(z)\;=\;\Ch_{-}(qz) + (1-\frac{s^2}{z^2})^n (1-\frac{q^2}{s^2z^2})^n \Ch_{-}(q^{-1}z)\;.
\end{equation}

So far it is unclear why we need these transformations. The reason is the following:\
\begin{stat}
Equations (\ref{TCh1}) and (\ref{TCh2}) have unique series solutions 
\begin{equation}
\Ch_{+}(z)\;=\;1+\sum_{k=1}^\infty \Ch_{+,k} z^{2k}\;,\quad
\Ch_{-}(z)\;=\;1+\sum_{k=1}^\infty \Ch_{-,k} z^{-2k}
\end{equation}
with the infinite convergency radius for the arbitrary coefficients inside the polynomial $T(z)$, eq. (\ref{T2}).
\end{stat}
\noindent
\textbf{Proof.} Consider for instance equation (\ref{TCh1}). It can be identically rewritten in the matrix form, 
\begin{equation}
\left(\begin{array}{c}
\ds \Ch_{+}(q^{-1}z)\\
\\
\ds \Ch_{+}(z)
\end{array}
\right)\;=\;
\underbrace{
\left(\begin{array}{cc}
\ds (-z)^n T(z) & \ds -(1-s^2z^2)^n(1-q^2\frac{z^2}{s^2})^n\\
\\
\ds 1 & 0 \end{array}\right)}_{\mathcal{L}(z)}
\;
\left(\begin{array}{c}
\ds \Ch_{+}(z)\\
\\
\ds \Ch_{+}(qz)
\end{array}
\right)\;.
\end{equation}
Note, $\ds \mathcal{L}(0)=\left(\begin{array}{cc} 2 & -1 \\ 1 & 0\end{array}\right)$ is a Jordan cell with the property 
$\mathcal{L}(0) \left(\begin{array}{c} 1 \\ 1\end{array}\right)= \left(\begin{array}{c} 1 \\ 1\end{array}\right)$, 
so that the infinite matrix product 
\begin{equation}\label{Zhor}
\left(\begin{array}{c}
\ds \Ch_{+}(q^{-1}z)\\
\\
\ds \Ch_{+}(z)
\end{array}
\right)\;=\;\mathcal{L}(z)
\mathcal{L}(qz)
\mathcal{L}(q^2z)
\cdots 
\left(\begin{array}{c}
\ds 1\\
\\
\ds 1
\end{array}
\right)
\end{equation}
is well defined and it is convergent everywhere in $z\in\mathbb{C}$. \hfill $\blacksquare$

The next step is to understand what theta–functions enter the definition of $H_{\pm}$. 
But, before doing this, consider the Wronskian of $\Ch_{+}$ and $\Ch_{-}$:
\begin{equation}
\mathcal{W}_{\chi}(z)\;=\;\Ch_{+}(q^{-1}z)\Ch_{-}(z) - (-q)^n[\frac{z}{s}]^n [\frac{sz}{q}]^n \Ch_{+}(z)\Ch_{-}(q^{-1}z)\;,
\end{equation}
so that 
\begin{equation}\label{Wr}
\mathcal{W}_{\chi}(z)\;=\;(-z^2)^n\, \mathcal{W}_{\chi}(qz)\;.
\end{equation}
Since $\Ch_{\pm}(z)$ are uniquely defined, the zeros of $\mathcal{W}_{\chi}(z)$ are also uniquely defined in the rectangle of periods:
\begin{equation}\label{Wr0}
\mathcal{W}_{\chi}(z)\;=\;\textrm{const}\times\prod_{k=1}^n (\frac{z^2}{w_k^2},q^2\frac{w_k^2}{z^2};q^2)\;,\quad
\prod_{k=1}^n w_k\;=\;1\;.
\end{equation}
Consider next the ratio,
\begin{equation}\label{Cra}
\mathcal{C}(z)\;=\;\frac{H_{+}(z)}{H_{-}(z)}\;.
\end{equation}
Equation $\mathcal{W}_{\chi}(z)=0$ can be identically rewritten in the form
\begin{equation}\label{Wqz}
\mathcal{C}(qz)\;=\;\mathcal{C}(z)\;.
\end{equation}

\bigskip

Now everything is ready to define finally the theta-functions $V_{\pm}$ and to construct the desired solution of  (\ref{TQ4}). 
Define \footnote{Note, substituting (\ref{Hpm},\ref{Fp},\ref{Fm}) into (\ref{WrQ}), one obtains identically $\mathcal{W}_{\chi}(z)=V_{+}(-z)V_{-}(z)$.}
\begin{equation}
V_{-}(z)\;=\;\prod_{k=1}^n (\frac{z}{w_k},q\frac{w_k}{z};q)_\infty\;,\quad
V_{+}(z)\;=\;(-z)^{-n}V_{-}(z)\;.
\end{equation}
Thus we make the denominators of $H_{+}(z)$ and $H_{-}(z)$ to be the same. The required entire solution of (\ref{TQ4}) is thus given by the linear combination 
\begin{equation}
H(z)\;=\;H_{+}(z) - C H_{-}(z)\;,
\end{equation}
where the poles cancellation condition is the Gutzwiller-type equation \cite{GW} 
\begin{equation}
\mathcal{C}(z)\;=\;C\quad \forall\; z\;:\;\;V_{-}(z)\;=\;0\;.
\end{equation}
According to (\ref{Wqz}), the number of equations here is $n-1$, namely
\begin{equation}\label{BAE}
\mathcal{C}(w_j)\;=\;\mathcal{C}(w_k)\quad \forall j,k=1,\dots,n
\end{equation}
while the number of unknowns is also $n-1$ according to the condition (\ref{Wr0}). 
The equations (\ref{BAE}) have the meaning of the Bethe Ansatz equations where $w_k$ play the role of Bethe variables. The real unknowns, however, are the coefficients of the polynomial $T(z)$. This makes a numerical analysis of the equations (\ref{BAE}) rather difficult. However, the number of variables equals to the numer of equations and therefore the problem is well posed and therefore solvable. \hfill $\blacksquare$

\subsection{The ground state in the thermodynamical limit}

Function (\ref{Cra}) in details becomes
\begin{equation}
\mathcal{C}(z)\;=\;\frac{H_{+}(z)}{H_{-}(z)}\;=\;(-z)^n \,\frac{\Si(sz)^n}{\Si(s/z)^n}\,
\frac{\Ch_{+}(z)}{\Ch_{-}(z)}\;.
\end{equation}
What we expect after taking into account the results of \cite{FVS}:
\begin{equation}
\Ch_{+}(z)\;=\;\prod_{m=1}^\infty \prod_{k=1}^n (1-q^{2m} \frac{z^2}{w_{k,m}^2})\;,\quad
\Ch_{-}(z)\;=\;\prod_{m=1}^\infty \prod_{k=1}^n (1-q^{2m} \frac{w_{k,m}^2}{z^2})\;.
\end{equation}
where
\begin{equation}
\frac{w_k}{w_{k,m}}\;\to \; 1\quad \textrm{when}\quad m\to \infty
\end{equation}
As well, we expect that the set of Bethe variables $w_k$, $k=1,\dots, n$,  in the thermodynamic limit $n\to\infty$ form some distribution on the unit circle with some distribution density.
At the same time, the set of $w_{k,m}$ for fixed $m$ forms a distribution along some curve with its own density, but such contour can be smoothly deformed to the unit circle so that the density of $w_{k,m}$ smoothly become the density of $w_k$. With this \emph{essentially strong assumptions} we obtain the following limit for  
$\mathcal{C}(z)$:
\begin{equation}
\frac{1}{n}\log\mathcal{C}(z)\;=\; \log(-z) + \log \frac{\Si(sz)}{\Si(s/z)}+
\frac{1}{2\pi\ii} \oint \frac{dw}{w} P(w) \log\frac{\ds (q^2\frac{z^2}{w^2};q^2)_\infty}{\ds (q^2\frac{w^2}{z^2};q^2)_\infty}\;=\;0\;,
\end{equation}
Here the density $P(w)$ is defined by the limit 
\begin{equation}\label{dens}
\lim_{n\to\infty}\frac{n}{2\pi} (\varphi_{k+1}-\varphi_k) P(\EXP^{\varphi_k})\;=\;1\;,\quad
P(\EXP^{\ii\varphi})\geq 0\;,\quad 
\int_{-\pi}^{\pi} \frac{d\varphi}{2\pi} P(\EXP^{\ii\varphi})\;=\;1\;.
\end{equation}
There is the helpful observation:
\begin{equation}
\frac{(q^2x^2;q^2)_\infty}{(q^2x^{-2};q^2)_\infty}\;=\;-x^{-1}\frac{\Si(x^{-1})}{\Si(x)}\;.
\end{equation}
Then the equation for the density can be rewritten as 
\begin{equation}
\log\frac{\Si(sz)}{\Si(s/z)}
+\frac{1}{2\pi\ii}\oint \frac{dw}{w} P(w) \log \frac{\Si(w/z)}{\Si(z/w)}
\;=\;0
\end{equation}
This equation has the following solution:
\begin{equation}\label{P00}
P(w)\;=\;1\;+\;\sum_{m=1}^\infty \frac{s^m + (-q/s)^m}{1+(-q)^m} \, (w^m+w^{-m})\;.
\end{equation}
Expression (\ref{P00}) for the density of zeros in the thermodynamical limit gives some bell-shape peak near $\varphi=0$, as expected. 
\\

Expression (\ref{P00}) has been obtained under rather strong assumptions. Hopefully, one can verify it. The first 
verification approach is the so-called ``tropical limit'', the limit $q\to 0$. In this limit $\Ch_{\pm}(z)=1$, so that  
\begin{equation}\label{trop}
\mathcal{C}(w)\;=\;\left(-w\frac{\Si(sw)}{\Si(s/w)}\right)^n\;\to \; \left(\frac{s-w}{1-sw}\right)^n\;=\;-1\;,
\end{equation}
what gives the solution 
\begin{equation}\label{part}
\frac{s-w_k}{1-sw_k}\;=\;\EXP^{2\pi \ii (k+\hl)/n}\;.
\end{equation}
The auxiliary transfer matrix then is given by the following expression:
\begin{equation}
(-z)^n T(z)\;=\;(1-s^{2n}) \prod_{k=1}^n (1-\frac{z^2}{w_k^2}) + (1-s^2z^2)^n + (s^2-z^2)^n\;.
\end{equation}
Combining then (\ref{part}) and (\ref{dens}) in the limit $n\to\infty$, one obtains
\begin{equation}
P(w)\;=\;1+\frac{sw}{1-sw} + \frac{sw^{-1}}{1-sw^{-1}}\;,
\end{equation}
what explicitly corresponds to (\ref{P00}) in the limit $q\to 0$. What else could be mentioned in regard to the tropical limit. The Reader can see $-1$ in the right hand side of (\ref{trop}). Thus the ground state is symmetric as it should be according to the general ideas of the quantum mechanics.  As well, the tropical limit in its simple form reproduces the ground state only. In other words, parameter $-q$ plays the role of the absolute temperature, and the tropical limit corresponds to the freezing of the system. More general approach to the tropical expansion will be formulated in the next section.\\
\\

As the second consistency test, one can recalculate the partition function already known \cite{BS}. The partition function is the Eigenvalue of $\hat{Q}$ per one site:
\begin{equation}
\hat{Q}(z)\;\sim\; \frac{\Ch_{+}(z)}{\Si(s/z)^nV_{+}(z)}\;\sim\; \frac{\Ch_{-}(z)}{\Si(sz)^nV_{-}(z)}\;,
\end{equation}
so that 
\begin{equation}
\frac{1}{n}\log\hat{Q}(z)\;=\;
-\log\Si(sz) + \frac{1}{2\pi\ii} \oint \frac{dw}{w} \, P(w)\, \log\Si(\frac{z}{w})\;,
\end{equation}
and, correspondingly, the partition function per one site is 
\begin{equation}
\frac{1}{n} \log\hat{Q}(z)\;=\; \sum_{m=1}^\infty \frac{ (-q)^m (s^{-m}-s^m)(z^m+z^{-m})}{m(1-q^m)(1+(-q)^m)}\;.
\end{equation}
This expression reproduces the result from \cite{BS}. The same result can be directly obtained from the Wronskian relation (\ref{WrQ}) without any assumption about the structure if the Bethe Ansatz. Recall, the physical regime in our notations is  $\ds 0 < -q<sz,s/z<1$.

\section{Tropical expansion and the structure of the spectrum }

\subsection{Method of the tropical expansion }

The method of the tropical expansion developed here is similar to those developed in \cite{Master}.

For Reader's convenience we commence with the repetition of the equations to be solved. The first one is the Baxter equation (\ref{TH}):
\begin{equation}\label{TH1}
\ds T(z) H(z)  \; =\; (-)^n
(1-sz)^{n}(1+q\frac{z}{s})^{n}  H(qz) \;+\; 
(1-\frac{s}{z})^{n}(1+\frac{q}{sz})^{n} H(q^{-1}z)\;,
\end{equation} 
where $H(z)$ is entire. In addition, sometimes it is convenient to use the Wronskian equation (\ref{WrQ}), rewritten in the terms of the entire function $H$:
\begin{equation}\label{WrH}
\begin{array}{l}
\ds 
(1-\frac{z}{s})^n (1+\frac{q}{sz})^n H(q^{-1}z) H(-z) \,-\,
(1+\frac{z}{s})^n (1-\frac{q}{sz})^n H(z) H(-q^{-1}z)\;=\\
\\
\ds =\; W\times\left( h(\frac{z}{s})^n h(-sz)^n \,-\, h(-\frac{z}{s})^n h(sz)^n \right)
\end{array}
\end{equation}
where $W$ is a multiplier introduced since a normalisation of $H(z)$ is undefined. Also, $\ds h(z)\;=\;(z,q/z;q)_\infty$ is in fact Jacoby's theta-one.
\\

Since $H(z)$ is entire, it has the following structure:
\begin{equation}\label{Hq}
H(z)\;=\;\sum_{k\in\mathbb{Z}} z^k q^{Q_k} H_k(q)\;,
\end{equation}
where $Q_k$ for sufficiently big $k$ is some positive quadratic function. The coefficients $H_k(q)$ can also be decomposed with respect to $q$,
\begin{equation}
H_k(q)\;=\;\sum_{j=0}^\infty H_{k,j} q^j\;.
\end{equation}
Evidently, this form of $H(z)$ is nothing but the $q$ decomposition.  The limit $q\to 0$ is known as the tropical limit, so we call the $q$ -- decomposition as the tropical expansion. The Baxter equation (\ref{TH1}), where the coefficients of $T(z)$ also are some seria in $q$, is the subject of the tropical expansion, 
\\

To construct the solutions to (\ref{TH1}) as the expansion in $q$, we have to define firstly the ``initial condition'', or ``seeding state''. Such seeding state would define completely the whole $q$ expansion. We will not give the complete $q$ series in this paper, we will give only the ``seeds''.
\\

\subsection{Equations for the tropical expansion}

For the beginning we collect all equations that can be used for the construction of the seeds. We could use polynomials $H_0(z)$, $H_0(qz)$ and $H_0(q^{-1}z)$ representing $H(z)$, $H(qz)$ and $H(q^{-1}z)$ in the leading order of $q$ when $q\to 0$. They are different polynomials related however by a gluing principle. For instance, if
\begin{equation}
H_0(z)=a z + b z^{-1}\;,
\end{equation}
then,
\begin{equation}
H_0(q^{-1}z)\;=\;q^{-1} ( az+a'z^2+\cdots)
\end{equation}
where $a',\dots$ appear from the elements of explicit $H(z)$ which disappear in the definition of 
 $H_0(z)$. Coefficient $a$ is the same in $H_0(z)$ and in $H_0(q^{-1}z)$. This simple observation we call the gluing principle.

When $H_0(z),H_0(q^{\pm 1}z)$ are defined, then the required seed is defined:
\begin{equation}
H(z) \;=\; az + bz^{-1} + qa'z + qb'z^{-1} + \cdots
\end{equation}
The main observation following from all these considerations is that for the seeding we need to construct three different polynomials related by the gluing principle.
\\

Now we can collect the equations defining the seeds. They are the Baxter equation (\ref{TH1}):
\begin{equation}\label{TH2}
\ds T_0(z) H_0(z)  \; =\; (-)^n
(1-sz)^{n}  H_0(qz) \;+\; 
(1-\frac{s}{z})^{n} H_0(q^{-1}z)\;,
\end{equation} 
and the Wronskian:
\begin{equation}\label{WrH2}
\begin{array}{l}
\ds 
(1-\frac{z}{s})^n H_0(q^{-1}z) H_0(-z) \,-\,
(1+\frac{z}{s})^n H_0(z) H_0(-q^{-1}z)\;=\\
\\
\ds =\; W_0\times\left((1-\frac{z}{s})^n (1+sz)^n \,-\, (1+\frac{z}{s})^n (1-sz)^n \right)
\end{array}
\end{equation}
A similar equation can be written for the shift $z\to qz$. Remarkably, the Wronskian relations in the tropical limit can be ``solved'':
\begin{equation}\label{HH0}
\left\{
\begin{array}{l}
\ds 
H_0(q^{-1}z) H_0(-z) \;=\; W_0 \times \left( (1+sz)^n - (s+z)^n K(z) \right)\;,\\
\\
\ds H_0(qz) H_0(-z) \;=\; W_0 \times \left( (1+sz^{-1})^n - (s+z^{-1})^n K'(z) \right)\;,
\end{array}\right.
\end{equation}
where $K(z)=K(-z)$ и $K'(z)=K'(-z)$ -- arbitrary symmetrical functions.
\\

In addition, we will need the holomorphic functions $\Ch_{\pm}$ introduced in the previous section.
Since we will look for excited states, we need more general expression for the auxiliary transfer matrix (\ref{T2}):
\begin{equation}\label{T3}
T(z)\;=\; (v+v^{-1})\, z^n \, + \,  \cdots \, + \,  (v+v^{-1})\, (-z)^{-n}\;,
\end{equation}
where
\begin{equation}
v\;=\;\vop_1\vop_2\cdots\vop_n\;.
\end{equation} 
For the model considered in this paper, $v$ is an unitary number. Note that for $v\neq 1$ the holomorphic approach requires an essential revision since the matrix product (\ref{Zhor}) is not well defined.

However, there is another model \cite{BS}, a dual one, for which the Baxter equation as well as the Wronskian equation remain the same, but the spectrum of $v$ becomes discrete:
\begin{equation}\label{vm}
v\;=\;q^{^m\!\!/\!_2}\;,\quad m\in\mathbb{Z}\;.
\end{equation}
This dual model has in particular another physical regime, $q>0$, so that $q^{\hl}$ is acceptable.
There is a temptation to call $m$ the complete spin. The formulas for the holomorphic functions from the previous section require a minimal change:
\begin{equation}\label{H2}
H_{+}(z)\;=\;\frac{(s/z,-q/sz;q)_\infty^n}{V(z)}z^{n+^m\!\!/\!_2}\Ch_{+}(z)\;,\quad
H_{-}(z)\;=\;\frac{(sz,-qz/s;q)_\infty^n}{V(z)}z^{-^m\!\!/\!_2}\Ch_{-}(z)\;,
\end{equation}
where 
\begin{equation}
V(z)\;=\;\prod_{k=1}^n (\frac{z}{w_k},q\frac{w_k}{z};q)_\infty\;,\quad
\prod_{k=1}^n w_k\;=\;\pm 1\;,
\end{equation}
and the holomorphic solutions are defined by 
\begin{equation}\label{ch2}
\begin{array}{l}
\ds \Ch_{+}(q^{-1}z)\;=\;(-z)^n q^{^m\!\!/\!_2} T(z)\Ch_{+}(z) - q^m (1-s^2z^2)^n(1-q^2z^2/s^2)^n\Ch_{+}(qz)\;,\\
\\
\ds \Ch_{-}(qz)\;=\;z^{-n} q^{^m\!\!/\!_2} T(z) \Ch_{-}(z) - q^m (1-s^2/z^2)^n (1-q^2/s^2z^2)^n \Ch_{-}(q^{-1}z)\;.
\end{array}
\end{equation}
These solutions are well defined for $m\geq 0$, and moreover, the matrix products in (\ref{Zhor}) are absolutely convergent for $m>0$. The change in the Wronskian is also minimal:
\begin{equation}\label{twr}
\mathcal{W}_\chi(z)\;=\;
\Ch_{+}(q^{-1}z)\Ch_{-}(z) - q^{n+m} [z/s]^n [q/sz]^n \Ch_{+}(z)\Ch_{-}(q^{-1}z)
\;\sim\; V(z)V(-z)\;.
\end{equation}

\subsection{Ground states}

We start the construction of the seeds with the ground state for general unitary $v$, what includes the case $m=0$. This seed is described uniquely by the trivial condition $H_0(z)\;=\;1$. 
Looking at (\ref{TH2}), one concludes that $H_0(q^{-1}z)$ is a polynomial of the positive power $n$, while $H_0(qz)$ is a polynomial of the negative power $n$. Looking at (\ref{HH0}), one concludes that $K,K'$ are constants. Taking then into account the gluing condition, one deduces
\begin{equation}\label{Hpm2}
H_0(q^{-1}z) \;=\;
\frac{(1+sz)^n - (s+z)^n K}{1-s^n K}\;,\quad
H_0(qz)\;=\;
\frac{(1+sz^{-1})-(s+z^{-1})K'}{1-s^n K'}\;.
\end{equation}
Using these $H_0(q^{\pm 1} z)$ in the Baxter equation (\ref{TH2}) and inspecting the powers  $z^n$ and $(-z)^{-n}$ in order to fix $K,K'$, one obtains
\begin{equation}
K\;=\;K'\;=\;-\frac{v}{1-s^nv} - \frac{v^{-1}}{1-s^nv^{-1}}\;.
\end{equation}
Thus, the seed in constructed uniquely for the arbitrary unitary $v$.

It could be helpful to obtain all leading coefficients in the decomposition of (\ref{Hq}). 
It is not difficult to do considering equation (\ref{TH1}) with $z\to q^k z$. Omitting details, we give the final answer:
\begin{equation}
\begin{array}{l}
\ds H_0(q^{-k}z)\;=\;
q^{-n k(k-1)/2} z^{n(k-1)} \left( \frac{v^k-v^{-k}}{v-v^{-1}} H_0(q^{-1}z) - 
\frac{v^{k-1}-v^{1-k}}{v-v^{-1}} (s+z)^n\right)\;,\\
\\
\ds H_0(q^kz)\;=\;
q^{-n k(k-1)/2} z^{-n(k-1)} \left( \frac{v^k-v^{-k}}{v-v^{-1}} H_0(qz) - 
\frac{v^{k-1}-v^{1-k}}{v-v^{-1}} (s+z^{-1})^n\right)\;.
\end{array}
\end{equation}
These formulas satisfy the gluing principle. They define the whole perimeter of the decomposition  (\ref{Hq}). It can be seen as a boundary condition for the perturbation theory in $q$ being understood as a sort of Cauchy problem. Having the boundary condition defined, the whole series in $q$ can be elementary constructed since in each new order of $q$ the coefficients of decomposition are the subject of a over-defined system of linear equations.

\subsection{Excitations for $m>0$.}

It worth to note ones more, the excitations for $m>0$ are not defined for the model considered.
The aim of this subsection is a sort of logical completeness, to demonstrate ``one-partical states'' before we turn to ``two-particle states'' in the next subsection.

In order to understand what happens for $v=q^{^m\!\!/\!_2}$ and $m>0$, we suggest to inspect the holomorphic equations (\ref{ch2}).  The second term in the right hand side there disappear in the limit $q\to 0$ for $m>0$.
Thus, taking $\Ch_{\pm}(z)=1$, we obtain for the seed
\begin{equation}
\Ch_{+}(q^{-1}z) = (-z)^n q^{^m\!\!/\!_2} T_0(z) =\prod_{k=1}^n (1-\frac{z^2}{w_k^2}),\;\;\;
\Ch_{-}(qz) = z^{-n} q^{^m\!\!/\!_2} T_0(z) =\prod_{k=1}^n (1-\frac{w_k^2}{z^2}),
\end{equation}
where
\begin{equation}
\prod_{k=1}^N w_k \; = \; 1\;,
\end{equation}
what can be established by inspection of the outmost elements of $T_0(z)$. The Wronskian  (\ref{twr}) is simple, 
\begin{equation}
\mathcal{W}_{\chi}(z)\;=\;\Ch_{+}(q^{-1}z)\;,\quad \textrm{so that }\quad
V_0(z)\;=\;\prod_{k=1}^n (1-\frac{z}{w_k})\;.
\end{equation}
Now we can construct $H_0(z)$:
\begin{equation}\label{H0m}
H_0(z)\;=\;H_{+}(z) + C H_{-}(z)\;=\;z^{-^m\!\!/\!_2} \,
\frac{z^m (z-s)^n + C (1-sz)^n}{V_0(z)}\;.
\end{equation}
As the result, we come to a simple quantisation condition: the coefficient $C$ must be chosen so that $V_0(z)$ cancel out. In addition,
\begin{equation}\label{H1m}
H_0(q^{-1}z)\;=\;q^{-^m\!\!/\!_2} (-)^n z^{^m\!\!/\!_2}\prod_{k=1}^n (1+\frac{z}{w_k})\;,\quad
H_0(qz)\;=\;C q^{-^m\!\!/\!_2} z^{-^m\!\!/\!_2} \prod_{k=1}^n (1+\frac{w_k}{z})\;.
\end{equation}

We can demonstrate how this condition works in the case of a ``one-particle states''  $m=1$. It is expected in this case,
\begin{equation}\label{PC}
P(z,C)\;=\;z(z-s)^n + C (1-sz)^n \;=\; C \prod_{k=0}^n (1-\frac{z}{w_k})\;,\quad
\prod_{k=0}^n w_k \;=\;(-)^{n+1} C\;.
\end{equation}
Since $\prod_{k=1}^n w_k\;=\;1$, we obtain
\begin{equation}
w_0\;=\;(-)^{n+1}C\;.
\end{equation}
The condition for $C$ is then 
\begin{equation}
P((-)^{n+1}C,C)\;=\;0\;,
\end{equation}
and solving it, we obtain
\begin{equation}\label{Cj}
C\;=\;C_j\;=\;(-)^{n+1} \frac{1-\omega^j s}{s-\omega^j}\;,\quad \omega\;=\;\EXP^{2\pi\ii/n}\;.
\end{equation}
This expression demonstrates a one-particle state with the momentum $\omega^j$, as expected. The resulting seed state is 
\begin{equation}
H_0(z)=(-)^{n+1}z^{-\hl}(\frac{1-\omega^j s}{s-\omega^j} - z)\;,
\end{equation}
and the values of $H_0(q^{\pm 1}z)$ are given by (\ref{H1m}), where (\ref{PC}) and (\ref{Cj}) are to be taken into account.

\subsection{Even excitations for $m=0$.}

This subsection is devoted to the construction of seeds in the sector of unitary $v$, what includes the case $m=0$ of the dual model. The result can be interpreted as an ``even number of particles''.

Since the holomorphic approach does not work, we will consider the Baxter equation (\ref{TH2}) 
and the Wronskian equations (\ref{HH0}).

The logic is the following. Presumably, $H_0(z)$ is a Laurent polynomial of degree $m$,
\begin{equation}
H_0(z)=z^m + \cdots + z^{-m}\;,
\end{equation}
where all coefficients are undefined. What could be a form of $H_0(q^{-1}z)$ in this case? Evidently,
\begin{equation}\label{HA1}
H_0(q^{-1}z)\;=\;q^{-m} z^m A_{+}(z)
\end{equation}
where $A_{+}(z)$ is a polynomial of some positive degree. Similarly,
\begin{equation}\label{HA2}
H_0(qz)\;=\;q^{-m}z^{-m} A_{-}(z)\;,
\end{equation}
where $A_{-}(z)$ is a polynomial of some negative degree. 

Next, we are looking for excited states in the sector $v\in\mathbb{T}$, so that the $z^{\pm n}$ elements in $T_0(z)$ become neglectable in all orders $q^{-m}$, $m>0$, so that the Baxter equation  (\ref{TH2}) provides
\begin{equation}\label{Adeg}
\textrm{deg} A \leq n-2\;.
\end{equation}
We take this observation into account and turn to the degrees in the Wronskian equation  (\ref{HH0}):
\begin{equation}
\textrm{deg}A+2m=N+\textrm{deg}K\;.
\end{equation}
It follows, under condition (\ref{Adeg}) and at least when $m=1$, one obtains
\begin{equation}
\textrm{deg}K\;=\;0\;.
\end{equation}
Therefore, at least for $m=1$ the following scenario for excited states can be formulated:
we take
\begin{equation}
K,K' \;=\; \textrm{const}
\end{equation}
and try to solve (\ref{HH0}), which takes the following form:
\begin{equation}\label{HH2}
\begin{array}{l}
\ds A_{+}(z) z^m H_0(-z) \;=\; q^m W_0 \left( (1+sz)^n - K (s+z)^n\right)\;,\\
\\
\ds A_{-}(z) z^{-m} H_0(-z) \;=\; q^m W_0 \left( (1+sz^{-1})^n - K' (s+z^{-1})^n \right)\;.
\end{array}
\end{equation}
Since $H_0(-z)$ is the common multiplier, the resultant of the right hand sides must be zero. It gives the relation between $K$ and $K'$:
\begin{equation}
KK'=1\;.
\end{equation}
It is convenient to introduce parameter $k$ instead of $K,K'$ as follows:
\begin{equation}\label{KKk}
K\;=\;k^n\;,\quad K'=k^{-n}\;.
\end{equation}
More detailed analysis of the Wronskian relations allows one to fix parameter $k$.

Using (\ref{KKk}), one can rewrite the Wronskian relations as
\begin{equation}
\begin{array}{l}
\ds A_{+}(z) z^m H_0(-z) \;=\; q^m W_0 (1-k^ns^n) \prod_{j=0}^{n-1} \left( 1 + z \frac{s-k\omega^j}{1-ks\omega^j}\right)\;,\\
\\
\ds 
A_{-}(z)\;=\; -k^{-n} z^{2m-n} A_{+}(z)\;.
\end{array}
\end{equation}
where $\ds \omega\;=\;\EXP^{2\pi\ii/n}$. One can fix $H_0(-z)$ now. Namely, one can disjoin the set $\mathbb{Z}_n$ into two subsets:
\begin{equation}
\{0,1,2,\dots,n-1\}\;=\;I_{2m}^{} \cup I_{n-2m}
\end{equation}
where the number of the elements in $I_{2m}^{}$ equals to $2m$. We can fix then 
\begin{equation}\label{h03}
H_0(z) \;=\; (-z)^{-m}\prod_{j\in I_{2m}} \left( 1 - z \frac{s-k\omega^j}{1-ks\omega^j}\right)\;.
\end{equation}
This choice corresponds to the choice of momenta of $2m$ particles. This scheme works evidently only for $2m<N$.
As the result, $A_{\pm}$ become
\begin{equation}\label{AA}
\begin{array}{l}
\ds 
A_{+}(z)\;=\;q^m W_0 (1-k^ns^n) \prod_{j\in I_{n-2m}} \left( 1 + z \frac{s-k\omega^j}{1-ks\omega^j}\right)\;,\\
\\
\ds A_{-}(z)\;=\; -k^{-n}q^mW_0(1-k^ns^n) \prod_{j\in I_{n-2m}} \left(z^{-1}+\frac{s-k\omega^j}{1-ks\omega^j}\right)\;.
\end{array}
\end{equation}
The last effort is the gluing of $H_0$ and $A_{\pm}$. Namely,
\begin{equation}
H_0(z)\;=\;  (-z)^{-m} + \cdots + (-z)^m \prod_{j\in I_{2m}} \frac{s-k\omega^j}{1-ks\omega^j}
\end{equation}
and hence 
\begin{equation}
H_0(q^{-1}z)\;\sim\; q^{-m}z^m 
(-)^m \prod_{j\in I_{2m}} \frac{s-k\omega^j}{1-ks\omega^j} \;=\; q^{-m}z^m A_{+}(0)
\end{equation}
and
\begin{equation}
H_0(qz)\;\sim\;q^mz^{-m} (-)^m \;=\; q^mz^{-m} A_{-}(\infty)\;.
\end{equation}
Extracting $A_{+}(0)$ and $A_{-}(\infty)$, one obtains
\begin{equation}\label{skl3}
q^m W_0 (1-k^ns^n) \;=\; (-)^m \prod_{j\in I_{2m}} \frac{s-k\omega^j}{1-ks\omega^j}\;=\;
-(-)^m k^n \prod_{j\in I_{n-2m}} \frac{1-ks\omega^j}{s-k\omega^j}
\end{equation}
with the consistency condition 
\begin{equation}
k^{2n}\;=\;1\;.
\end{equation}
Two branches are obtained as the result,
\begin{equation}\label{whatk}
k\;=\;1\quad \textrm{or}\quad k\;=\;\omega^{\hl}
\end{equation}
they correspond to symmetric and antisymmetric wave functions from the point of view of Quantum Mechanics\footnote{$k=1$ gives antisymmetric wave function for $n=2$ while $k^2=-1$ gives the antisymmetric one.}. The resulting formula for $T_0(z)$ becomes
\begin{equation}
T_0(z)\;=\;q^{-m} (1-k^ns^n) \prod_{j\in I_{n-2m}} \left( z\frac{s-k\omega^j}{1-ks\omega^j} - z^{-1}\frac{1-ks\omega^j}{s-k\omega^j}\right)\;.
\end{equation}
As the result, the seed for $2m$-particles states in the regime $v\in\mathbb{T}$ is given now by (\ref{h03},\ref{HA1},\ref{HA2},\ref{AA},\ref{skl3}).
\\


\section{Conclusion}

What is done in this paper:
\begin{enumerate}
\item The Clebsh-Gordan basis for our particular representation is constructed.
\item The CG basis was used for construction of separated states, i.e. the Eigenstates for the off-diagonal element $\hat{B}(\lambda)$ of the monodromy matrix. 
\item Construction of the separated basis can be extended to e.g. Faddeev–Volkov  \cite{FV} and other models where $6j$ symbols are known as the quantum dilogarithms.
\item The Eigenstate of the model in the separated basis is the product of solutions of the Baxter equation. Traditionally we call the solution of Baxter equation Baxter's $\hat{Q}$ operator.
\item The Functional Bethe Ansatz approach fixes the class of functions $\hat{Q}$ operator. 
The procedure of construction of the solutions to Baxter equation in the required class of functions is formulated.
\item This procedure allows to construct solution of our analogue of the Bethe Ansatz for the ground state in the thermodynamical limit.
\item Thus, the eigenstate is constructed.
\item In addition, the Baxter equation is studied using the tropical expansion method. In addition to the ground state, the lowest elementary excitations are described and classified.
\end{enumerate}

\vspace{1cm}
\noindent
\textbf{Acknowledgements.} The author would like to thank R. Kashaev, V. Bazhanov and V. Mangazeev for valuable discussions. The work was partially supported by the ARC grant DP190103144.

\end{document}